# Silicon nitride C-band grating coupler with reduced waveguide back-reflection using adaptively corrected elliptical grates

Ibrahim Ghannam[1] [0000-0002-0952-6581], Florian Merget[1] [0000-0001-8208-1681] and Jeremy Witzens[1] [0000-0002-2896-7243]

[1] Institute of Integrated Photonics, RWTH Aachen University, 52074 Aachen, Germany
ighannam@iph.rwth-aachen.de

**Abstract.** We present experimental results for a fully etched C-band grating coupler with reduced back reflection fabricated in an 800 nm silicon nitride platform. Back-reflections are reduced by symmetrically interrupting the first few grates around the center axis of the propagating light. The span of the etched grates is gradually increased until they cover the full width. By interrupting the grates, light is reflected back obliquely, which leads to the excitation of higher-order modes that are scattered out of the structure. While this approach has been previously shown in silicon, it comes with a significant penalty in coupling efficiency of around 2.4 dB of extra loss in the layer stack investigated here. In this work, we present the design and measurement results of a grating coupler in which waveguide-to-waveguide back-reflections are suppressed by ~10 dB with this technique, while at the same time mitigating excess insertion losses by reshaping the grates as ellipses of varying eccentricity. This helps to compensate the phase front error induced by the interruption of the grates. This correction does not affect the level by which the back-reflection is suppressed, but reduces the insertion loss penalty from 2.4 dB to 1 dB.

**Keywords:** Light coupling, grating coupler, back-reflection suppression, silicon nitride.

## 1    Introduction

Grating couplers (GCs) enable dense photonic integrated circuits (PICs) by allowing coupling from a single-mode fiber to anywhere on the chip's surface. This has led to the wide adoption of GCs as a preferred method to couple light into the silicon-on-insulator (SOI) and silicon nitride (SiN) platforms [1, 2], particularly as it comes to individual device characterization or accessing taps inside a system. More recently, there has been a focus on optimizing GCs to reduce waveguide-to-waveguide back-reflections [3-6], since these can cause parasitic laser feedback that can otherwise limit the usability of GCs in systems such as external cavity lasers (ECLs) and transceivers [7-9].

In this work, we focus on reducing the back-reflection that occurs at the boundary between the input taper of a focusing GC and the first few grating trenches. This type of back-reflection is particularly high in fully-etched GCs, which is a requirement in some open access foundry processes. While partial etches and apodization allow for a substantial reduction in back-reflection, increased requirements as for example required to obtain ECL linewidths in the sub-kHz range [10, 11] warrant here too the introduction of advanced design concepts.

Here, we present a continuation of our work in [6], showing the experimental realization of a fully etched C-band SiN GC featuring both suppressed back-reflection and a very moderate insertion loss penalty that was proposed and modeled there.

The reduction in waveguide-to-waveguide back-reflection is achieved by altering the field profile of the back-reflected light and exciting higher-order modes that are not supported by the single-mode input waveguide. This is achieved by gradually modifying the span of the first few grates, as also previously shown in the SOI platform [4, 5]. However, this also distorts the phase front of the field propagating through the grating coupler, after emission from the single mode waveguide, so that significant excess losses are induced [6]. To mitigate this, the shape of the grates is adapted on an individual basis to compensate for the incurred phase error. This modification does not have a negative effect on the back-reflection reduction.

The devices are implemented in the 800 nm nitride thickness Ligentec platform. The reference GC, prior to any of these modifications, has an insertion loss of 7.8 dB. Modification of the grate span suppresses back-reflections by ~10 dB, but adds an additional 2.4 dB of losses. After modification of the grates into elliptical shapes with grate-dependent eccentricity, these excess losses drop to 1 dB.

## 2     Grating Coupler Design

The design of the reference GC is shown in Fig. 1(a). The device has a pitch of 1.17 μm with a fill factor of 44.87%, defined as the ratio of the width of the unetched section to the total pitch. The length of the input taper is 30 μm, spanning from the input waveguide to the onset of the grates. The taper's full angle is 25.6 degrees. The input waveguide only supports a single transverse electric (TE) ground mode and has a width of 600 nm. The light is coupled in and out through a fiber array with an 11° fiber polish and the GC is operated as a forward-coupling device. The reference device provides a baseline against which to compare the performance of the modified devices.

Figure 1b, 1c shows the designs modified for reducing the back-reflection. In both devices, the span of the trenches in the first three grates, that are interrupted in the center of the beam propagating inside the slab, is increased gradually. The first trench covers an angle ϕ on both sides of the beam, the second trench an angle 2ϕ, and so on until the trenches fill the entire span. This essentially forms a segmented boundary that reflects light at an odd angle that cannot reenter the single mode waveguide.

Device B has an identical pitch and fill factor to device A, so that the out-coupling angle is essentially maintained. However, the missing grates in the center of the device result in a perturbation of the phase front, that is inherited by the emitted beam and increases the insertion losses. A second excess loss mechanism results from the scattered field simply missing in the center region of the truncated grates. While the latter cannot be helped without compromising the mechanism leading to the back-reflection reduction, the phase front distortion can be accommodated by modifying the grate shape accordingly.

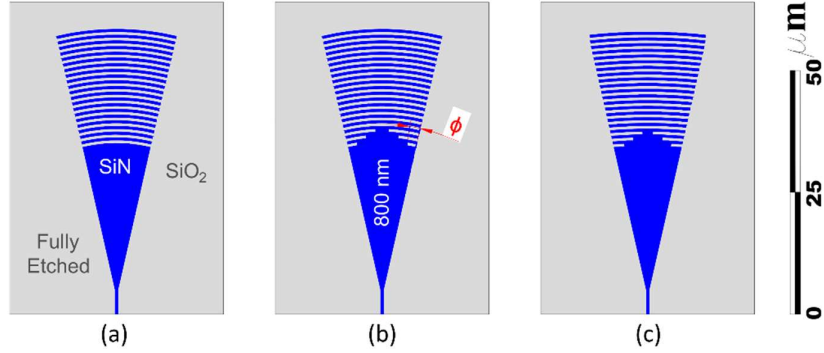

**Fig. 1.** Layouts of (a) the reference GC with circular grates ("Device A"), (b) the modified GC with three interrupted grates but otherwise identical geometry ("Device B"), and (c) the modified GC in which the grates are further converted into ellipses of varying eccentricity ("Device C"). In b and c, the span of the trenches in the first three grates is controlled by the angle φ and the grate index. The trench spans, from the edge of the taper, over an angle given by the grate index multiplied by φ (on both sides). φ is set to 3.65 degrees in both b and c.

Figure 2 illustrates the design flow by which the modified grate eccentricity is obtained. The correction is performed in two steps. First, the grates are segmented into seven angular sectors (2 times the number of interrupted grates + 1) as shown in Fig. 2(a). In each sector, grates are moved towards the input beam by an increment $d$ for each left-out grate. This displacement $d$ is calculated using the equation

$$\frac{2\pi n_{sl}}{\lambda_0} d = \frac{2\pi(n_{sl}-n_{ox})}{\lambda_0} w + \frac{2\pi n_{ox}}{\lambda_0} sin(\theta) d \qquad (1)$$

with $n_{sl}$ the effective index of the SiN slab (1.88), $\lambda_0$ the free-space wavelength (1550 nm), $n_{ox}$ the refractive index of $SiO_2$ (1.45), $w$ the width of the trenches (645 nm), and $\theta$ the angle of the scattered beam in $SiO_2$ (11°). After algebraic manipulation, the equation can be rewritten as

$$d = \frac{n_{sl}-n_{ox}}{n_{sl}-n_{ox} sin(\theta)} w \qquad (2)$$

This modification is targeted to scatter the field with the correct phase for obtaining a beam emission with a flat free-space wave-front. However, the abrupt transitions at the interfaces between the angular sectors leads to scattering of the field, so that smoothing of the resulting geometry is required.

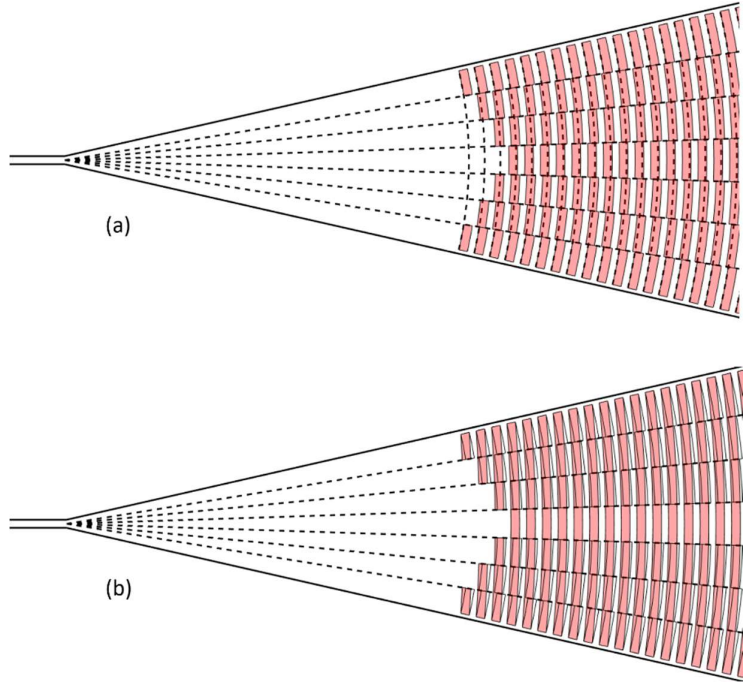

**Fig. 2.** (a) First approximation of the modification required for accommodating the phase front distortion. Grates in each sector are shifted towards the direction of the input waveguide to compensate for the increased phase accretion in the angular sectors with unetched regions. The dashed curves correspond to the original grates (identical in device B and the reference GC) and are plotted to visualize the displacement. (b) Grates after smoothing by fitting to an ellipse. Here, the outline of the segmented grates is overlaid for comparison.

In a second step, the segmented grates are replaced by elliptical grates that best match the segments. This is done by modifying the eccentricity of the ellipses, which vary between 0.85 and 0.64 for the first 20 grates. This also allows the device to be in compliance with fabrication restrictions. The final geometry is shown in Fig. 2(b).

A series of simulations were performed to illustrate the devices' functionality further. Fig. 3 shows the intensity of the back reflected field as it reaches the interface between the input waveguide and the taper. The data is generated using 3D finite difference time domain (FDTD) simulations. A packet of light centered around 1550 nm is launched from a mode source into the waveguide on the left side of Fig. 3. The field is recorded with a monitor that takes a snapshot of the intensity along the center plane of the SiN film at a time at which the reflected field reaches again the input waveguide. For the reference device, most of the light is reflected back into the ground mode of the input waveguide. For devices B and C, light is reflected obliquely due to the modification of the first three grates. This leads to the excitation of higher-order modes that are not supported by the input waveguide and couple out, reducing the intensity of the waveguide back-reflection.

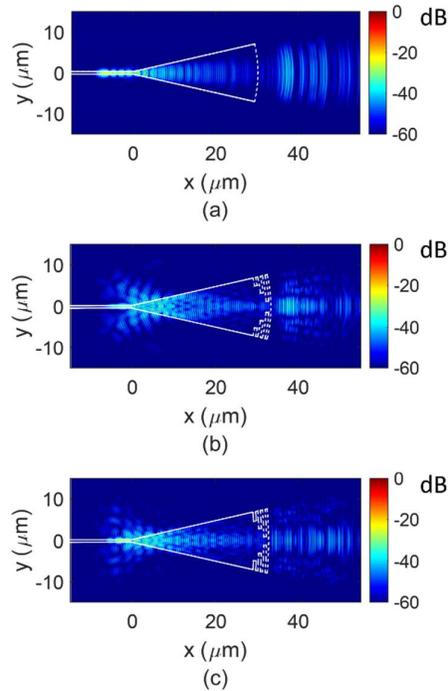

**Fig. 3.** Intensity of back-reflected light for (a) the reference device, (b) device B, and (c) device C, recorded at the center of the SiN film from an FDTD simulation at a time step at which the back-reflected light has again reached the input waveguide (identical time step for all panels).

## 3   Measurement Results

The devices were characterized and evaluated in a similar way to [6]. Test structures consist of two GCs connected by a waveguide loop with a 184 µm length. One GC was used as an input, and the other as output. A tunable laser source was used to obtain the wavelength dependent transfer function. Besides the insertion losses, the waveguide-to-waveguide back-reflection can be extracted by analyzing the ripples in the device response, wherein Fourier filtering is used to reject ripples arising from reflections occurring elsewhere in the setup.

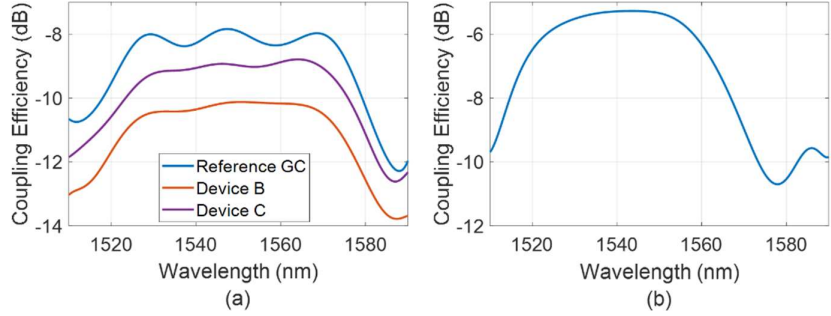

**Fig. 4.** (a) Smoothed coupling efficiency of the three devices. (b) Smoothed coupling efficiency of the reference device from an earlier run.

The smoothed coupling efficiency of a single GC is shown in Fig. 4(a) for all three devices. The losses recorded from the loops were divided by two to extract the insertion loss for a single GC. The best coupling efficiency of the reference device and device B are -7.8 dB and -10.2 dB, respectively, obtained at a wavelength of 1550 nm. Device C has a coupling efficiency of -8.8 dB at a wavelength of 1564 nm. The 3-dB bandwidth is around 65 nm for all three devices. The large ripples that can be seen in Fig. 4(a) correspond to a length that matches that of the GC input taper (30 µm), which indicates multiple reflections between its interfaces. These can already be seen to be less prominent in the modified GCs.

In an earlier run, the reference device and device B had better insertion losses of -5.4 dB and -8.1 dB, respectively, 2.1-2.4 dB better than here. The cause of this drop in coupling efficiency is still under investigation, but is likely due to a change in the under-cladding and over-cladding layer thicknesses. However, the relative improvement in the performance of device C relative to device B matches expectations very well. The data from previous runs gives reason to believe that the insertion losses of device C can be improved to ~ -6.4 dB by reverting to the previous process conditions.

Fourier filtering is used to extract the ripples with the wavelength periodicity corresponding to the loop length. Considering a path length of 248 µm including the waveguide path and tapers and a group index of 2.1, we expect the ripples' free spectral range (FSR) to be 2.3 nm. The optical spectra are taken with a 1 pm resolution (80,000 points between 1510 and 1590 nm). For such spectra, an FSR of 2.3 nm corresponds to a spectral peak at the data point of index 35 after applying a discrete Fourier transform (DFT), which can be observed in Fig. 5(b). It can be seen that this peak is very significantly attenuated for devices B and C. The second peak is caused by the measurement setup, as it was present in different measurements at the same data point index regardless of the length of the loop on the SiN chip. The peak around index 35 is filtered by removing the background spectrum and transformed back with an inverse Fourier transform. This results in the ripples shown in Fig. 5(c). To convert these into a power back-reflection coefficient $R$, the peak-to-peak ripple values are used in the following equation [6, 12]

$$R = \frac{\sqrt{ER} - 1}{\sqrt{ER} + 1} \tag{3}$$

where *ER* stands for extinction ratio and corresponds to the ratio between the maximum and minimum transmission recorded in the transfer function of the GC loop (i.e., before dividing the losses by two). The ripples represented in Fig. 5(c) are from the transfer function of the whole loop and can be directly plugged into Eq. (3).

The back-reflection spectrum for the different devices is shown in Fig. 5(d). Both Devices B and C achieve significant reduction in back reflection compared to the reference device, in the range of ~10 dB over a bandwidth of 60 nm that includes the C-band. In addition, it can be observed that device C outperforms device B even in terms of back-reflection suppression, which is attributed to the fact that more of the light is coupled out.

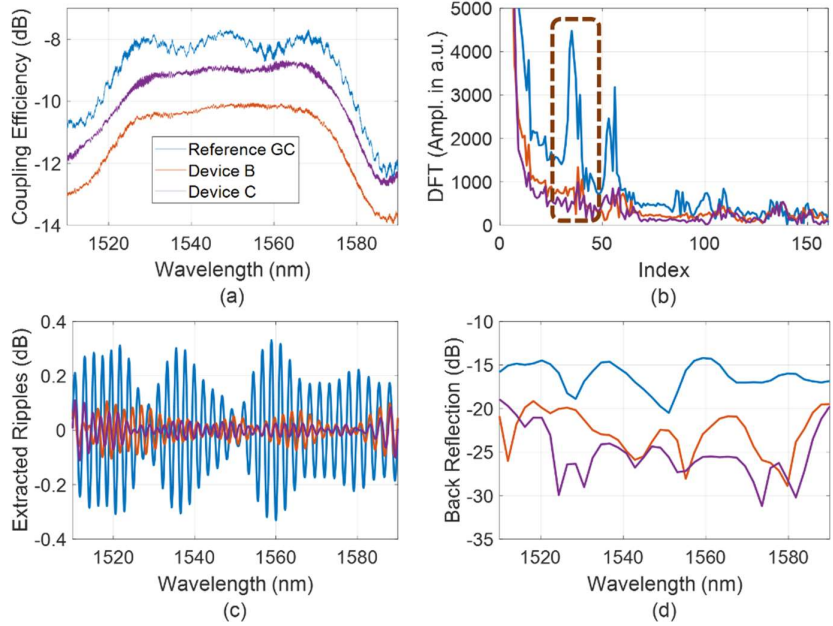

**Fig. 5.** (a) Raw coupling efficiency of a single GC, (b) DFT applied to the GC transfer function in dB, (c) extracted ripples after filtering and the inverse DFT, with results normalized for the whole loop (i.e., multiplied back by two relative to panel (a)), and (d) extracted back-reflection (color coding identical in all panels).

## 4  Conclusion

We present a fully etched SiN grating coupler with a reduced back-reflection based on truncated grates. The reduction in back-reflection usually comes at the expense of a reduced coupling efficiency. To counter this and reduce the insertion loss caused by the truncation of the grates, we adapt the grate geometry in a way that compensates for phase errors. The new grating coupler has elliptical grates of varying eccentricity. ~10 dB of back-reflection suppression is obtained across the 60 nm 3-dB bandwidth while reducing excess insertion losses from 2.4 dB to 1 dB.

**References**

1. Marchetti, R., Lacava, C., Carroll, L., Gradkowski, K., and Minzioni, P.: Coupling strategies for silicon photonics integrated chips. Photon. Res. 7, 201–239 (2019).
2. Romero-García, S., Merget, F., Zhong, F., Finkelstein, H., and Witzens, J.: Silicon nitride CMOS-compatible platforms for integrated photonics applications at visible wavelengths. Opt. Expr. 21(12), 14036–14046 (2013).
3. Benedikovic, D., Alonso-Ramos, C., Pérez-Galacho, D., Guerber, S., Vakarin, V., Marcaud, G., Le Roux, X., Cassan, E., Marris-Morini, D., Cheben, P., Boeuf, F., Baudot, C., and Vivien, L.: L-shaped fiber-chip grating couplers with high directionality and low reflectivity fabricated with deep-UV lithography. Opt. Lett. 42, 3439–3442 (2017).
4. Song, J. H., Snyder, B., Lodewijks, K., Jansen, R., and Rottenberg, X.: Grating coupler design for reduced backreflections. IEEE Photon. Technol. Lett. 30(2), 217–220 (2018).
5. Zou, J., Zhang, Y., Hu, J., Wang, C., Zhang, M., and Le, Z.: Grating coupler with reduced back reflection using $\lambda/4$ offset at its grating sub-teeth. J. Lightwave Technol. 37, 1195–1199 (2019).
6. Ghannam, I., Merget, F., and Witzens, J.: Silicon nitride C-band grating couplers with reduced waveguide back-reflection. In Proc. SPIE 12007, Optical Interconnects XXII, 120070N (2022).
7. Ghannam, I., Shen, B., Merget, F., and Witzens, J.: Silicon Nitride External Cavity Laser with Alignment Tolerant Multi-Mode RSOA-to-PIC Interface. J. Sel. Top. Quant. Electron. 28(1), 1501710 (2022).
8. Tran, M. A. *et al.*: Extending the spectrum of fully integrated photonics. arXiv:2112.02923 (2021).
9. Shang, C., Wan, Y., Selvidge, J., Hughes, E., Herrick, R., Mukherjee, K., Duan, J., Grillot, F., Chow, W. W., and Bowers, J. E.: Perspectives on Advances in Quantum Dot Lasers and Integration with Si Photonic Integrated Circuits. ACS Photon. 8(9), 2555–2566 (2021).
10. Xiang, C., Morton, P. A., and Bowers, J. E.: Ultra-narrow linewidth laser based on a semiconductor gain chip and extended $Si_3N_4$ Bragg grating. Opt. Lett. 44(15), 3825–3828 (2019).
11. Maier, P. *et al.*: Sub-kHz-Linewidth External-Cavity Laser (ECL) with $Si_3N_4$ Resonator used as a Tunable Pump for a Kerr Frequency Comb. J. Lightwave Technol. 41(11), 3479–3490 (2023).
12. Shen, B., Lin, H., Sharif Azadeh, S., Nojic, J., Kang, M., Merget, F., Richardson, K. A., Hu, J., and Witzens, J.: Reconfigurable frequency-selective resonance splitting in chalcogenide microring resonators. ACS Photon. 7(2), 499–511 (2020).